\documentclass[pra,twocolumn,floatfix,aps]{revtex4}

\usepackage{gensymb}
\usepackage{amsmath}
\usepackage{makeidx}
\usepackage{amsmath}
\usepackage{graphicx}
\usepackage{ulem}
  
\usepackage[usenames]{color}
\usepackage{hyperref}
\usepackage[T1]{fontenc} 
\begin{document}

\title{Dynamically stable  two- and four-droplet  solitons in a very strongly dipolar NaCs   condensate}


 \author{S. K. Adhikari\footnote{sk.adhikari@unesp.br}}


\affiliation{Instituto de F\'{\i}sica Te\'orica, UNESP - Universidade Estadual Paulista, 01.140-070 S\~ao Paulo, S\~ao Paulo, Brazil}

\begin{abstract}

A strongly dipolar Bose-Einstein condensate (BEC) of Dy atoms could sustain different types of 
states not possible in a non-dipolar BEC.  Motivated by the observation of a very strongly
dipolar condensate of NaCs molecules [N. Bigagli  et al.,   Nature 631, 289 (2024)], with dipolar interaction stronger by more than an order of magnitude compared to that of Dy atoms, 
we demonstrate that in a very strongly dipolar NaCs BEC, 
it is possible to have a two- and a four-droplet metastable soliton, axially free along the polarization $z$ direction. 
 In this study we   employ imaginary-time propagation of  an improved mean-field model,
including the Lee-Huang-Yang interaction.  
The  dipolar solitons are  subject to  an expulsive Gaussian potential at the center and a harmonic potential $-$ both acting in the $x$-$y$ plane. The phase-coherent solitons, possessing the isolated droplets, are free to move along the  $z$ axis without any relative motion between the droplets.
The dynamical stability and mobility of these solitons are demonstrated  by real-time propagation employing the converged imaginary-time wave function as the initial state.

\end{abstract}


\maketitle

\section{Introduction}

 The observation of a dipolar Bose-Einstein condensate (BEC) of $^{52}$Cr \cite{dipbec, y2, y3, y4, y5, y6}, $^{166}$Er \cite{y7}, $^{168}$Er
\cite{y8}, $^{164}$Dy \cite{y9, y10, y11, y12} atoms with large magnetic dipole moments
 opened a new avenue of research. 
More recently, a very strongly dipolar 
BEC of NaCs molecules was observed \cite{NaCs}. 
The dipolar length of Dy atoms, which is a measure
of the strength of the dipolar interaction, viz. Eq. (\ref{eq.dl}), has the value    $a_{\mathrm{dd}}= 130.8a_0$, where $a_0$ is the Bohr radius.
However, the effective electrical dipole moment of a NaCs molecule can be  controlled 
by the microwave-shielding technique \cite{Bohn},  which allows to have a NaCs molecule of  very large 
dipolar length in  the range  $ a_{\mathrm{dd}}= 1000a_0$  to $25000a_0$ \cite{NaCs}, much larger than the same of Dy  atoms.

The most remarkable phenomenon
in a strongly dipolar harmonically trapped BEC  is the
observation of a single droplet \cite{y13, 2d3} 
of size much
smaller than the harmonic oscillator trap length, in addition to a self-bound droplet in free space \cite{y13,santos,x22}. With
the increase of the number of atoms in the harmonically trapped dipolar BEC,
multiple droplets are formed  \cite{y13,y15,y16,y17,y18}.

Another important research area in BEC is the statics  and dynamics of quantum solitons. A soliton or solitary wave is an one-dimensional  (1D) self-reinforcing wave packet that maintains its shape  due to a cancellation of nonlinear attraction and dispersive repulsion,
while it propagates at a constant velocity \cite{x1,x2}. { The 1D soliton is analytic and obeys strict conservation laws (energy, momentum),  and maintains its shape after a collision.} 
Solitons
have been studied in water wave, Bose-Einstein condensate (BEC) \cite{x1} and a general nonlinear medium \cite{nlm} including nonlinear 
optics \cite{x2}. 
 In three dimensions  (3D), the realization of such a propagating soliton is possible by confining a BEC 
 in the $x$-$y$ plane by  a trap
and allowing it to propagate freely along the untrapped $z$ direction \cite{perez},  thus creating a quasi-one-dimensional (quasi-1D) configuration.   
A quasi-1D  soliton was predicted
\cite{perez} for attractive interaction  and created in a BEC of $^7$Li \cite{x4,x5} and $^{85}$Rb 
\cite{x6} atoms. 
There have also been many studies of quasi-1D solitons in dipolar BECs under different conditions
\cite{c30,c31,c32,c33}.  {    Quasi-1D solitons were also  found \cite{so1,so2,so3} in a spin-orbit (SO) coupled BEC equilibrated by an external trap  and SO-coupling interaction.   }  
In addition,  quasi-1D  solitons are also possible \cite{c27}
in a
dipolar BEC.
There are numerous examples of similar quasi-1D solitons in a BEC 
\cite{as1,as2,xyz,as3,as4,as5,as6,xyz,as8}.   
  { Nevertheless, all quasi-1D solitons are formed in the presence of a trap in the $x$-$y$ plane
and hence are numeric and only obey the conservation laws approximately.}

In this paper, we address the interesting question ``In view of the multiple droplet formation in a trapped dipolar BEC \cite{y15,y16,y17,y18,y19},  is it possible to have a multiple-droplet 
quasi-1D  soliton in a strongly dipolar BEC?''   To this end, we demonstrate that it is indeed possible to have  a metastable quasi-1D two- or a four-droplet soliton in a very strongly dipolar  BEC of NaCs molecules  free to move along the polarization $z$ direction.  
 The quasi-1D trap  constitutes of a harmonic one in the $x$-$y$ plane  and an expulsive  Gaussian potential at the center of the same plane with no trap in the $z$ direction.    The  two- and the four-droplet  
 solitons are phase coherent and can move as a unified whole along the $z$ axis
without any relative motion between the spatially-separated droplets.
{
These quasi-1D solitons, like all quasi-1D BEC solitons, 
stabilized  by the combined effect of the confining trap and the long-range dipolar interaction, are numeric  in nature and  are not analytic solitons in the strict
mathematical sense and the term soliton is being used more loosely here to
highlight their stability and mobility.}
The energy of the two-droplet soliton is smaller than that of the four-droplet soliton 
and  the lowest-energy stable soliton is  an one-droplet soliton. As the dipolar interaction is reduced, by decreasing the number of molecules or the effective electric dipole moment of a molecule, the  droplets of the metastable soliton join together to form a  stable soliton of a single  droplet.

In theoretical investigations,  employing the mean-field  Gross-Pitaevskii (GP) model,
a trapped    dipolar BEC collapses in 3D for a strong dipolar interaction beyond a critical value \cite{c2,y6,c4,c5,c6}.
However, when an improved  Lee-Huang-Yang \cite{lhy} (LHY) interaction, appropriate for a dipolar system   \cite{qf1,qf2,qf3}, is included in this model, the collapse can be stopped \cite{santos,drop3}. 
As the number of atoms (or molecules)   is gradually increased in a trapped dipolar BEC in this model, 
a stable  droplet,  
and eventually, multiple droplets
are formed \cite{2d3,y13,blakie}.   
Using  this improved mean-field model,  we show that with an appropriate confining trap in the $x$-$y$ plane alone a very strongly dipolar BEC of NaCs molecules can sustain a  two- and a four-droplet soliton. { The self-bound quantum droplets  in free space  \cite{santos,x22}  are  mobile in all directions,  droplets in a harmonically trapped dipolar BEC \cite{y13,2d3} are immobile, 
 whereas the present droplet solitons are mobile only along the polarization $z$ direction. }

    In Sec. \ref{II} we present the improved mean-field model for the  very strongly dipolar BEC of NaCs molecules 
    with  repulsive contact and long-range dipolar interactions, including an  appropriate LHY  interaction. 
      In Sec. \ref{III}  we display numerical  results for the  formation of the metastable two- and four-droplet solitons. 
       The stationary states of the metastable solitons will be obtained by imaginary-time propagation and their 
       propagation dynamics  
       studied by real-time propagation.    In Sec. \ref{IV}  we present a brief summary of the present investigation.

\section{Improved Mean-field model}

\label{II}

We consider a  BEC of $N$ dipolar  molecules,  polarized along the $z$ axis, 
interacting through the following 
molecular  dipolar and  contact   interactions   \cite{dipbec,dip,yuka}
\begin{align}
V({\bf R})= &
\frac{ \mathrm{d_{eff}}^2}{4\pi\epsilon_0}U_{\mathrm{dd}}({\bf  R})
+\frac{4\pi \hbar^2 a}{m}\delta({\bf  r-r' }), 
\label{eq.con_dipInter} \\
U_{\mathrm{dd}}({\bf R}) =& \frac{1-3\cos^2 \theta}{|{\bf  R}|^3}, \label{dippot2}
\end{align}
where $m$ is the mass of a molecule, $a$ is the intermolecular $s$ wave scattering length, 
$\mathrm{d_{eff}}$ is the effective electric dipole moment of each NaCs molecule,  $\epsilon_0$ is the
permittivity of vacuum.  Here, $\bf r \equiv \{x,y,z\}$ and $\bf r' \equiv \{x',y',z'\}$ are the position vectors  of the  dipolar molecules
and $\theta$ is the angle made by  the relative position vector  $\bf R\equiv r-r'$ with the  polarization
$z$ direction.   The following dipolar length $a_{\mathrm{dd}}$ 
determines the  strength of dipolar 
interaction 
\begin{align}
a_{\mathrm{dd}}=\frac{ m\mathrm{d_{eff}}^2 }{ 12\pi \hbar ^2\epsilon_0},
 \label{eq.dl}
 \end{align}
 whereas the scattering length $a$ determines the  strength of contact interaction.
 
In this paper we employ  an  improved mean-field model   incorporating the LHY interaction \cite{lhy} appropriately modified for dipolar atoms and molecules \cite{qf1,qf2}. 
The formation of a two- or  a four-droplet soliton  mobile along the polarization $z$ direction    is described by the following  3D  GP equation including the  LHY interaction \cite{dipbec,dip,2d4,blakie,yuka}
\begin{align}\label{eq.GP3d}
 \mbox i \hbar \frac{\partial \psi({\bf r},t)}{\partial t} &=\
{\Big [}  -\frac{\hbar^2}{2m}\nabla^2
+U({\bf r})
+ \frac{4\pi \hbar^2}{m}{a} N \vert \psi({\bf r},t) \vert^2 \nonumber\\
&\ +\frac{3\hbar^2}{m}a_{\mathrm{dd}}  N
\int U_{\mathrm{dd}}({\bf R})
\vert\psi({\mathbf r'},t)\vert^2 d{\mathbf r}'  
\nonumber \\
& +\frac{\gamma_{\mathrm{LHY}}\hbar^2}{m}N^{3/2}
|\psi({\mathbf r},t)|^3
\Big] \psi({\bf r},t),\\
U({\bf r})&=\frac{m\omega^2}{2}(x^2+y^2) + V_0 \exp\left[-\frac{x^2+y^2}{\delta^2}\right] \label{potx} ,
\end{align}
where 
the trapping potential $U({\bf r}) $ is a combination of a weak confining  harmonic  and an expulsive   Gaussian potential, both in the $x$-$y$ plane and the dipolar system is free to move along the polarization $z$ direction; $\omega$  is the angular frequency of the  harmonic trap,    and $V_0$ and $\delta$ are the strength and width of the Gaussian potential, respectively. 
  The repulsive Gaussian potential keeps the droplets of the soliton apart  and allows the formation of a metastable two- or a four-droplet dipolar soliton in the $x$-$y$ plane with long extension in the $z$ direction.  
The wave function is  normalized as $\int \vert \psi({\bf r},t) \vert^2 d{\bf r}=1.$  The coefficient 
of the  LHY interaction $\gamma_{\mathrm{LHY}}$ in Eq. (\ref{eq.GP3d})
is given by \cite{qf1,qf2,blakie}
\begin{align}\label{qf}
\gamma_{\mathrm{LHY}}= \frac{128}{3}\sqrt{\pi a^5} Q_5(\varepsilon_{\mathrm{dd}}), \quad \varepsilon_{\mathrm{dd}}= \frac{a_{\mathrm{dd}}}{a},
\end{align}
where  the auxiliary function $ Q_5(\varepsilon_{\mathrm{dd}})$ is given by 
$ Q_5(\varepsilon_{\mathrm{dd}})=\ \int_0^1 dx(1-\varepsilon_{\mathrm{dd}}+3x^2\varepsilon_{\mathrm{dd}})^{5/2}. $
This function  is evaluated as \cite{blakie}
\begin{align}\label{exa} 
Q_5(\varepsilon_{\mathrm{dd}}) &=\
\frac{(3\varepsilon_{\mathrm{dd}})^{5/2}}{48}  \Re \left[(8+26\eta+33\eta^2)\sqrt{1+\eta}\right.\nonumber\\
& + \left.
\ 15\eta^3 \mathrm{ln} \left( \frac{1+\sqrt{1+\eta}}{\sqrt{\eta}}\right)  \right], \quad  \eta = \frac{1-\varepsilon_{\mathrm{dd}}}{3\varepsilon_{\mathrm{dd}}},
\end{align}
where $\Re$ denotes the real part.  Expression (\ref{exa}) for $Q_5(\varepsilon_{\mathrm{dd}})$  has been used in the numerical calculation of this paper.

We can rewrite 
Eq. (\ref{eq.GP3d})  in
the following  dimensionless form 
by scaling lengths in units of $l_0 = \sqrt{\hbar/m\omega_0}$,    time in units of $t_0=\omega_0^{-1}$,  angular frequency $\omega$ in units of $\omega_0$,  energy and $V$ in units of $\hbar\omega_0$,
and density $|\psi|^2$ in units of $l_0^{-3}$, where $\omega_0$ is a reference angular frequency:
\begin{align}\label{GP3d2}
\mbox i \frac{\partial \psi({\bf r},t)}{\partial t} & =
{\Big [}  -\frac{1}{2}\nabla^2
+U({\bf r}) + 4\pi{a} N \vert \psi({\bf r},t) \vert^2
\nonumber   \\ 
&+3a_{\mathrm{dd}}  N
\int 
U_{\mathrm{dd}}({\bf R})
\vert\psi({\mathbf r'},t)\vert^2 d{\mathbf r}'   \nonumber \\
&+\gamma_{\mathrm{LHY}}N^{3/2}
|\psi({\mathbf r},t)|^3  
\Big] \psi({\bf r},t),\\  
U({\bf r})&= \frac{1}{2}\omega^2\left(x^2+ y^2\right) +V_0  \exp\left[-\frac{x^2+y^2}{\delta^2}    \right] ,
\label{pot}
\end{align}
where, and in the following, without any risk of confusion, we will be using the same symbols to denote the new scaled variables. 
The normalization in the scaled variables remains unchanged:  $\int |\psi({\bf r,}t)|^2 =1.$
 
We can also derive  Eq. (\ref{GP3d2})   from the variational rule
$i {\partial \psi}/{\partial t} = {\delta E}/{\delta \psi^*}$
where $E$  is an energy   functional (energy per molecule of a stationary state) given by
\begin{align} \label{energy}
E &= \int d{\bf r} \Big[ {\frac{1}{2}|\nabla\psi({\bf r})|^2} +  U({\bf r})|\psi({\bf r})|^2\nonumber  \\
&+ \frac{3}{2}a_{\mathrm{dd}}N|\psi({\bf r})|^2 
\left. \int U_{\mathrm{dd}}({\bf R})
|\psi({\bf r'})|^2 d {\bf r'} \right. \nonumber \\
& + 2\pi Na |\psi({\bf r})|^4 +\frac{2\gamma_{\mathrm{LHY}}}{5} N^{3/2}
|\psi({\bf r})|^5\Big].
\end{align}

\section{Numerical Results}

\label{III}

To obtain the very strongly dipolar two- and four-droplet   solitons of NaCs molecules, 
we  solve   partial differential   GP  
 equation (\ref{GP3d2}),  numerically, using FORTRAN/C programs \cite{dip,yuka1} or their open-multiprocessing versions \cite{omp,ompF}, 
employing  the split-time-step Crank-Nicolson  
method using the imaginary-time propagation rule  \cite{crank}.  
 It is difficult to treat numerically  the divergent $1/|{\bf R}|^3$ term in the dipolar potential (\ref{dippot2})   in configuration space. To circumvent  this problem, this term    in  the nonlocal dipolar interaction integral in Eq.  (\ref{GP3d2}) is evaluated in the momentum $\bf k$
 space by a Fourier transformation \cite{dip}. 
 After  the problem is solved  in  the momentum space,  the desired solution in configuration space is obtained by a backward Fourier transformation.

  For the appearance of a quasi-1D two- or a four-droplet soliton we need a very strongly dipolar   BEC of NaCs molecules  with 
  {$a_{\mathrm{dd}}\gg a$}   \cite{2d3}.  In this study we take $a_{\mathrm{dd}} = 2000a_0$ and  $a=100a_0$, which makes the system very strongly dipolar. Although we are using NaCs molecules in this study, we do not include the effect of microwave shielding, which is beyond the scope of this investigation. 
For NaCs molecules    $m$(NaCs)    $\approx 156 \times 1.66054\times 10^{-27}$ kg; 
 we take the reference frequency $\omega_0 = 2\pi \times 180 $ Hz, consequently, the unit of length $l_0 =\sqrt{\hbar/m\omega_0}= 0.600$  $\mu$m, unit of time $t_0=\omega_0^{-1}=0.88$ ms.   The trap parameters in Eq. (\ref{potx}) will be taken as
 $\omega =2\pi \times 54$ Hz, $V_0/h = 1800$ Hz, $\delta = \sqrt{2}l_0 \approx 0.8485$ $\mu$m.  Here $l_0$ and $\omega_0$ are 
scaling parameters used in writing the dimensionless equations and have no effect on the final result.

      Although metastable, these solitons can be obtained by imaginary-time propagation, as they possess distinct spatial symmetry properties; an imaginary-time propagation scheme usually preserves the spatial symmetry of the initial state. 
   The metastable  two- and four-droplet solitons are dynamically stable and this dynamics can be studied by real-time propagation.  Given a push,
 these solitons are able to travel along the polarization $z$ direction without any visible deformation with a  uniform velocity, as will be demonstrated.

\begin{figure}[t!]  
\begin{center}
\includegraphics[width=\linewidth]{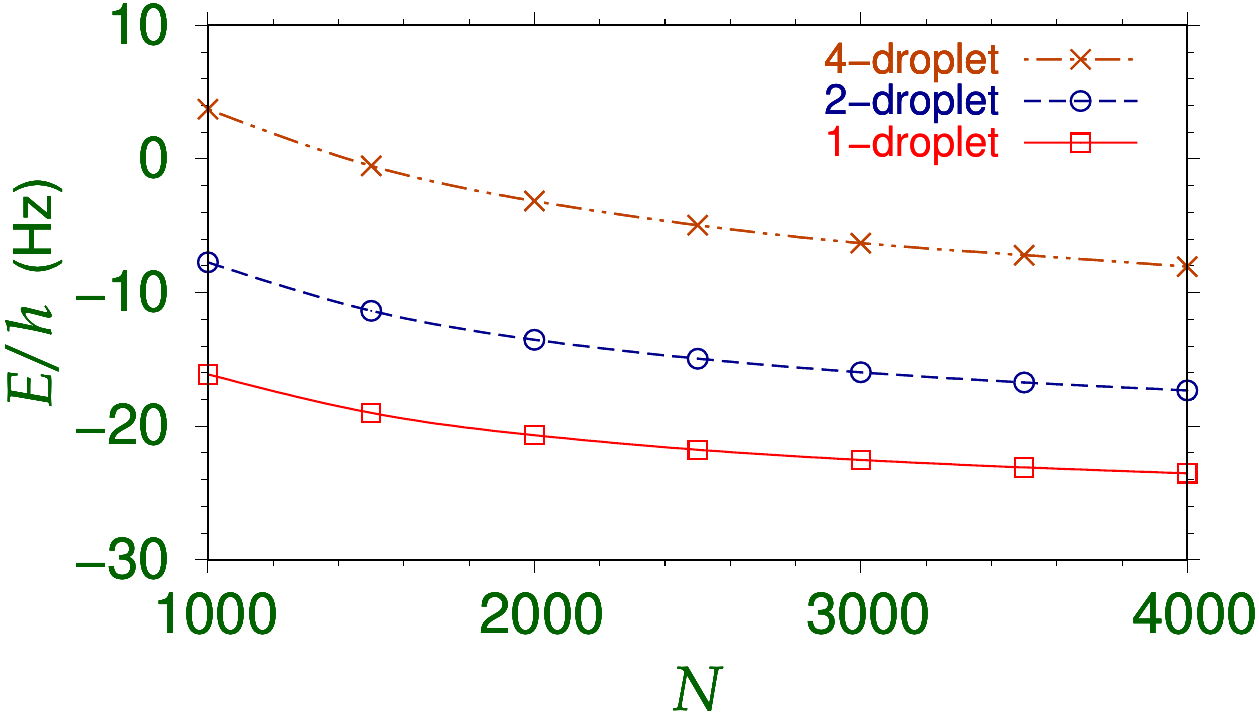}

\caption{Energy per atom $E/h$ in Hz of a four-droplet, a two-droplet and an one-droplet soliton for different number 
of NaCs molecules $N$. The points are the numerical results, which are joined by the lines to guide the eye. The dipolar length and the scattering length are $a_{\mathrm{dd}}=2000a_0, a=100a_0,$ respectively.   The trap parameters, viz. Eq. (\ref{potx}), are $\omega =2\pi \times 54$ Hz, $V_0/h = 1800$ Hz, $\delta = \sqrt{2}l_0,$ $ l_0=0.6$ $\mu$m.  These parameters are the same in all figures.}

\label{fig1}   
\end{center}
\end{figure}




To obtain a four-droplet soliton easily by the imaginary-time propagation routine, we took the initial state as four  droplets  placed at the corners of a square,  whose center is at the origin, for example, at $\{x,y\}=\{\pm \alpha,\pm \alpha\}.$  To obtain a two-droplet state the two droplets in the initial state are placed at $\{x,y\}=
 \{\pm \alpha,0\}.$  Because of the distinct spatial symmetry of a four- and a two-droplet soliton,  these states are remarkably stable and  a higher-energy   four-droplet soliton cannot easily decay to a lower-energy two-droplet soliton,  provided we employ  an accurate numerical scheme with small time and space steps with a large number of space discretization points.  In an approximate numerical scheme, this may not be true and a four-droplet initial state may lead  to a two-droplet, or, even,  an one-droplet soliton.  We take the space discretization steps along $x,y,z$ directions as $dx=dy=0.1, dz=0.125$, respectively, which is adequate for our study.   The employed time steps were $dt =0.1 (dx\times dy\times dz)^{2/3}$ in imaginary-time propagation and 
 $dt =0.025 (dx\times dy\times dz)^{2/3}$ in real-time propagation.  The number of space discretization points in $x$ and $y$ directions were $N_x=N_y=257$ and that in the $z$ direction could be as large as $N_z=769$.
  
 \begin{figure*}[htbp]
\begin{center}
\includegraphics[width=\linewidth]{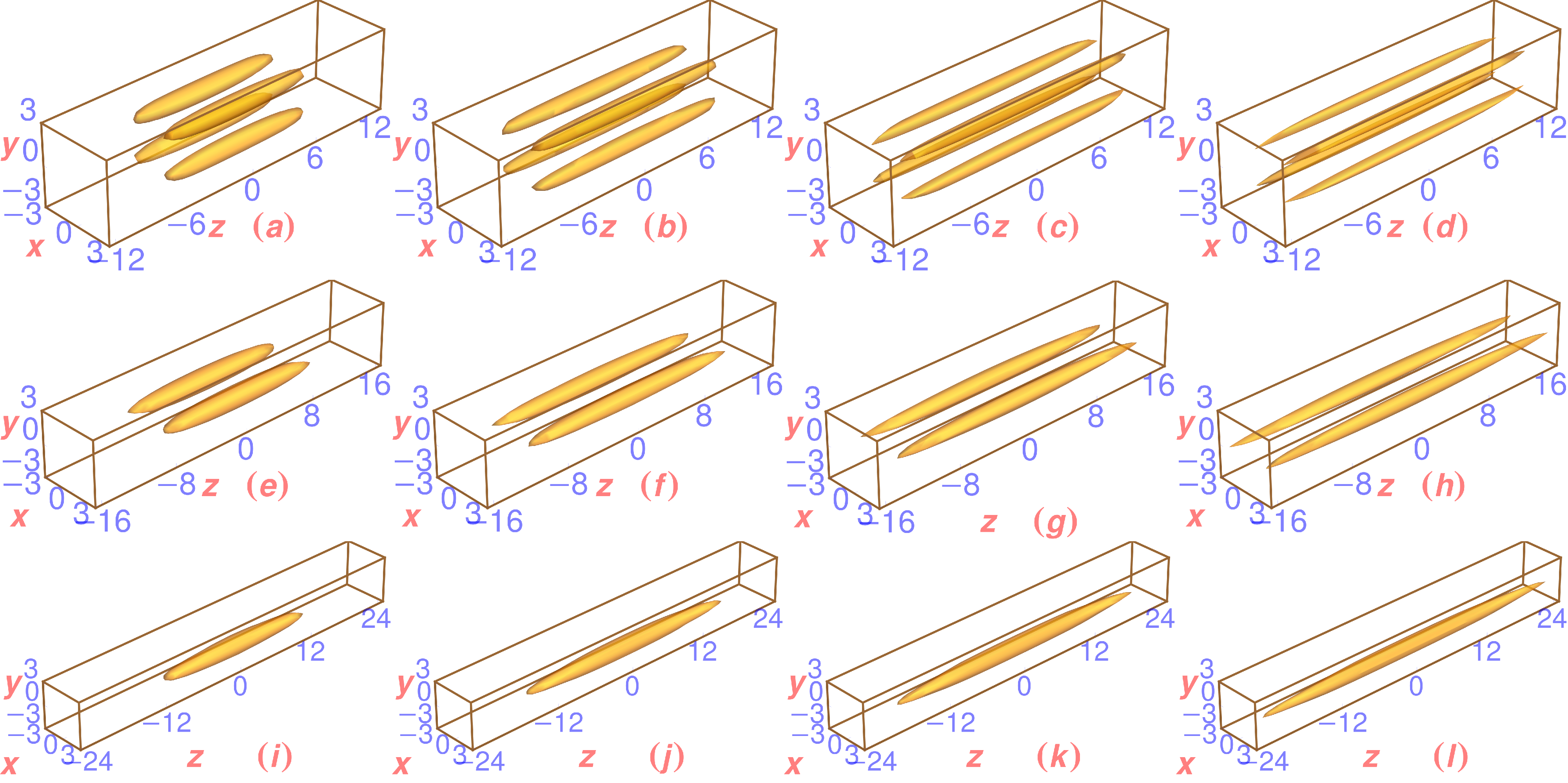} 

\caption{Isodensity plot of normalized density $|\psi(x,y,z)|^2$ ($\int |\psi({\bf r})|^2 d{\bf r}=1$) of a four-droplet soliton of $N=$  (a) 1000, (b) 2000, 
(c) 3000, and (d) 4000 NaCs molecules,  of a two-droplet soliton of  $N=$  (e) 1000, (f) 2000, 
(g) 3000, and (h) 4000 molecules, and  of an one-droplet soliton of $N=$  (i) 1000, (j) 2000, (k) 3000, and  (l) 4000   molecules. 
 The unit of lengths is $\mu$m. The density on contour is 
$\rho_{\mathrm{cont}}=4.6 \times 10^9$ cm$^{-3}$.  As the net density of molecules is $N|\psi(x,y,z)|^2$, the density of NaCs molecules  on contour in these cases are $N \rho_{\mathrm{cont}}$.
}  
\label{fig3} 
\end{center}
\end{figure*}

 We find that the solitons can be found for a relatively small number of molecules.
 In Fig. \ref{fig1} we display the energy per molecule (\ref{energy})  $E/h$ for the four-, two-, and the one-droplet solitons for different number of  molecules $N$  from 1000 to 4000.  For a fixed number of molecules, the energy  $E/h$ of the four-droplet (two-droplet) soliton is larger than that of the two-droplet (one-droplet) soliton. The energy $E/h$, as well as  the total energy of the soliton  $NE/h$,   for a fixed number of droplets, 
 decreases as the number of molecules increases. For the present set of parameters, in imaginary-time propagation we could not stabilize  a three-droplet  soliton or a soliton with more than four droplets. In imaginary-time propagation, 
these states eventually decay to a two-droplet or an one-droplet soliton. 
 {  For a different  set of interaction and trap parameters,  it might be possible to stabilize the solitons with a different number  of droplets, which could be a study of future interest.
The repulsive dipolar interaction in the $x$-$y$ plane and the expulsive Gaussian potential at the center 
keep the droplets apart while the  harmonic trap  binds  them in a metastable state  to form a two-droplet or a four-droplet soliton.   However, if we take a more complex trap in the $x$-$y$ plane, consisting of multiple barriers, preliminary study indicates that the  number of droplets in the soliton will increase, which is beyond the scope of this investigation.  The interaction between solitons and droplets depends on the
relative phase.  In the present case of phase-coherent solitons the phase between the different droplets is zero.
 }

The confining potential (\ref{pot}) in the $x$-$y$ plane is one with ring topology.  In place of using the appropriately placed four or two droplets in the initial state in imaginary-time propagation, if we use a Gaussian function as the initial state, we obtain a highly metastable hollow cylindrical soliton state with ring topology.  This state  has higher energy than the four-, two-, and one-droplet solitons, viz. Fig. \ref{fig1}.   With a prolonged time propagation, the hollow cylindrical soliton  evolves into the four-droplet soliton (result not shown in this paper).  However, it might be possible to stabilize a  metastable hollow cylindrical soliton with  ring topology  for a different set of parametars, which could be a very interesting work of future investigation.

 The evolution of the few-droplet soliton for different number of droplets and different number of molecules is best illustrated through a isodensity plot of the solitons. In Fig. \ref{fig3} we show the evolution of a four-droplet soliton by the isodensity plot of normalized density $|\psi(x,y,z)|^2$ for (a) $N=1000$,
 (b) $N=2000,$  (c) $N=3000,$ and (d) $N=4000$ NaCs molecules.  As the number of molecules increases from Fig. \ref{fig3}(a) through (d), the system becomes increasingly dipolar and the length of the droplets along the polarization $z$ direction increases, although the spatial extension in the $x$-$y$ plane of the four-droplet soliton remains essentially the same for different $N$.  The spatial extension in the $x$-$y$ plane  
 is essentially controlled by the minimum of the confining trap (\ref{pot}) in that plane.
 If we had plotted the net density $N|\psi(x,y,z)|^2$, in place of normalized density $|\psi(x,y,z)|^2$, and used the same cut-off density in the plots, both the length of a droplet and its section in the $x$-$y$ plane will increase, as $N$ increases, as we will see in the following through a plot of the net integrated two-dimensional  (2D)  \cite{ajp} and 1D densities.  The same trend continues in  the case of a two-droplet  soliton  as shown in Fig.  \ref{fig3} for   (e) $N=1000$,
 (f) $N=2000,$  (g) $N=3000,$ and (h) $N=4000$ molecules  and an one-droplet soliton as shown in Fig. \ref{fig3} for  (i) $N=1000$, (j) $N=2000$,   (k) $N=3000$, and (l) $N=4000$  molecules. For the same number $N$ of molecules,  each droplet of a   two-droplet soliton in Fig. \ref{fig3} is longer along the $z$ direction than the same of a four-droplet soliton. This is because, for  a fixed $N$, each droplet of a two-droplet soliton has twice as many molecules when compared to the same of  a four-droplet soliton.  A larger number of molecules in each droplet of a two-droplet   soliton leads to a larger dipolar interaction responsible for generating a longer droplet.  From the isodensity plot of an one-droplet soliton of Fig. \ref{fig3} we find that these droplets are longer than  both a four-droplet and a two-droplet soliton because of the largest number of molecules in an one-droplet soliton.  The dipolar interaction in an one-droplet soliton is the largest, which makes the one-droplet soliton the longest along the polarization $z$ direction.




\begin{figure}[t!] 
\begin{center}
\includegraphics[width=\linewidth]{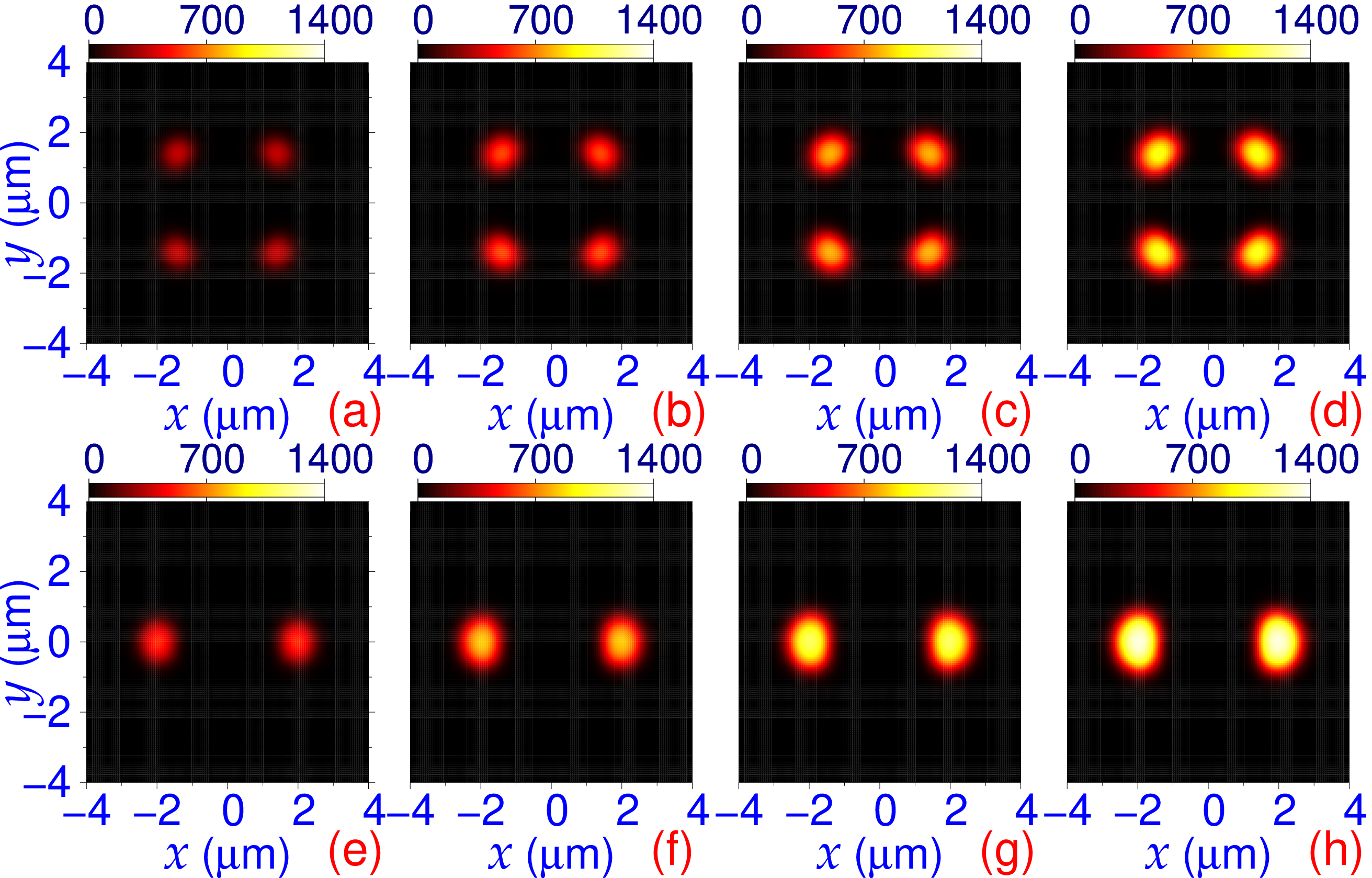} 

\caption{Contour   plot of net integrated 2D density $n(x,y) \equiv Nn_{\mathrm{2D}}(x,y)$ 
[$\int_{-\infty}^\infty dx \int_{-\infty}^\infty  dy n(x,y)=N$]
versus $\{x,y\}$ of four-droplet solitons  of Fig. \ref{fig3} for  $N=$  (a) 1000, (b) 2000, 
(c) 3000, and (d) 4000 NaCs molecules.  The same   of two-droplet solitons of Fig.  \ref{fig3} for  $N=$  (e) 1000, (f) 2000, 
(g) 3000, (h) 4000 NaCs molecules.
The unit of densities in the color box is $\mu$m$^{-2}$.
}

\label{fig5} 
\end{center}
\end{figure}

The different spatial extensions of these solitons can be studied employing the integrated 2D density in the $x$-$y$ plane $n_{\mathrm{2D}}(x,y)$ and the integrated 1D density  in the $z$ direction $n_{\mathrm{1D}}(z)$ defined, respectively, by
\begin{align}\label{2d}
n_{\mathrm{2D}}(x,y) &= \int_{-\infty}^{\infty} dz |\psi(x,y,z)|^2,\\
 n_{\mathrm{1D}}(z) &= \int_{-\infty}^{\infty} dx \int_{-\infty}^{\infty} dy  |\psi(x,y,z)|^2.
 \label{1d}
\end{align}
The integrated 2D density $n_{\mathrm{2D}}(x,y)$ is useful to study the spatial extension (or localization) of the soliton in the $x$-$y$ plane, the integrated 1D density $n_{\mathrm{1D}}(z)$ is useful to study the linear extension of the  same along the $z$ direction.

 \begin{figure}[t!]
\begin{center}
\includegraphics[width=.49\linewidth]{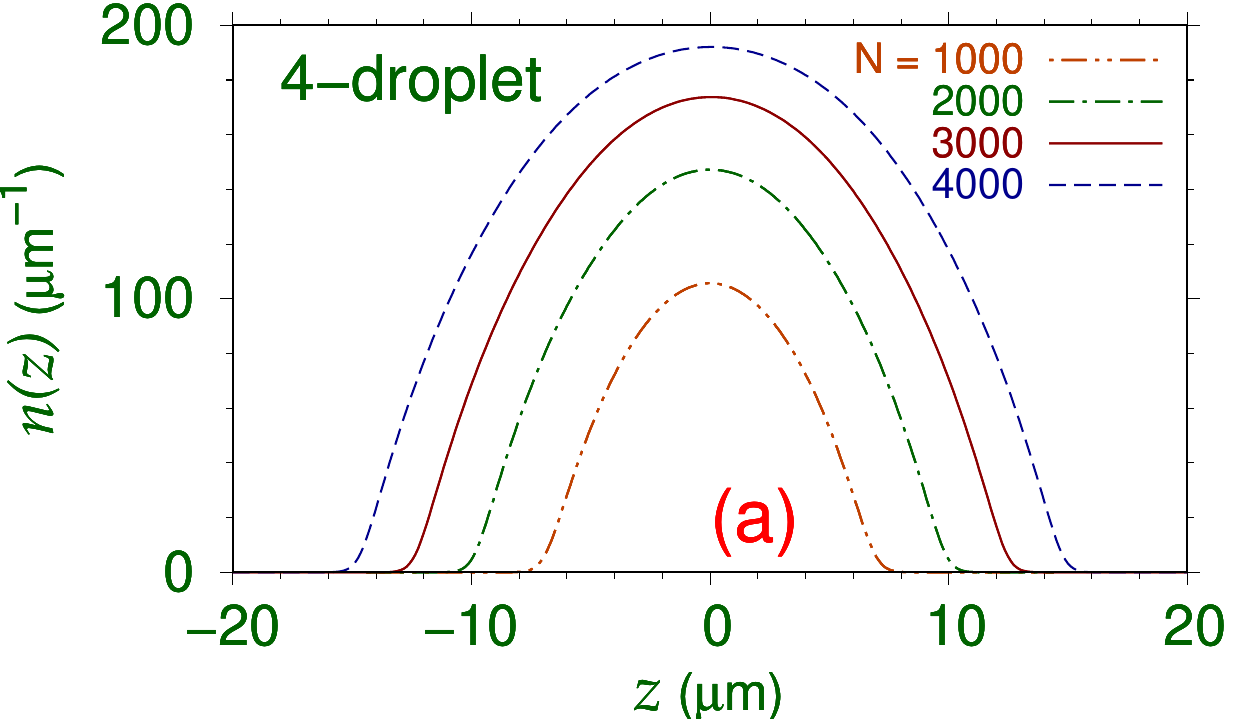} 
\includegraphics[width=.49\linewidth]{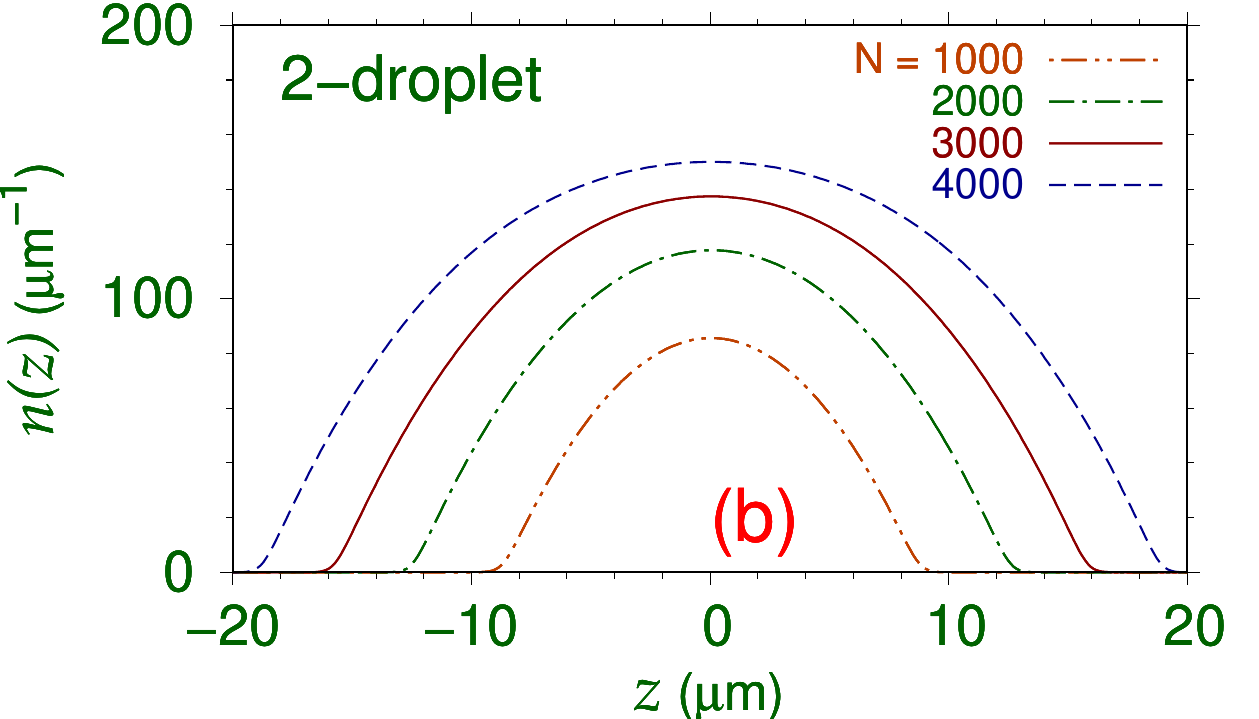}

\caption{Net integrated 1D density $n(z) \equiv Nn_{\mathrm{1D}}(z)$  [$\int_{-\infty}^\infty dz n(z)=N$] of (a) a four-droplet and  (b) a two-droplet  soliton  of Fig.  \ref{fig3}, for  $N=1000,2000,3000$, and 4000 NaCs molecules. 
}

\label{fig6} 
\end{center}
\end{figure}

In Fig. \ref{fig5} we illustrate the net integrated 2D density $Nn_{\mathrm{2D}}(x,y)$ through a contour plot of the same in the $x$-$y$ plane  of the four-droplet solitons  of Fig. \ref{fig3}    of  $N=$  (a) 1000, (b) 2000, 
(c) 3000, and (d) 4000 NaCs molecules.
 The same 
 of the two-droplet solitons  of Fig. \ref{fig3}   are shown in Fig. \ref{fig5} for   $N=$  (e) 1000, (f) 2000, 
(g) 3000, and (h) 4000 NaCs molecules.  As the number of molecules $N$ increases in plots in Figs. \ref{fig5}(a)-(d) of a four-droplet soliton, 
the  droplets not only become fatter,  with larger cross-section in the $x$-$y$ plane, but also have larger molecule density.  The same is true  about a two-droplet soliton shown in  Figs. \ref{fig5}(e)-(h).
However, the maximum molecule density and the cross section of a single droplet 
in a two-droplet soliton in    Figs. \ref{fig5}(e)-(h) are larger than the same in   
Figs. \ref{fig5}(a)-(d) in a four-droplet soliton, respectively for   $N=$   1000,  2000, 3000, and  4000, which is reasonable  as each droplet in a two-droplet soliton contains twice as many molecules as in a four-droplet soliton for a fixed $N$.

In Fig.\ref{fig6} we display the net integrated 1D density $Nn_{\mathrm{1D}}(z)$ through a plot of the same versus $z$ for (a) a four-droplet soliton and (b) a two-droplet soliton for $N=1000, 2000, 3000$  and 4000, which also reveals some relevant information.  In a four-droplet or a two-droplet soliton, the length along the $z$ direction increases with number of molecules $N$.  Also, for a fixed $N$, a two-droplet soliton is longer along the $z$ direction  than a four-droplet soliton as each droplet in the former has twice as many molecules when compared to the same in the latter.  

In a four-droplet or a two-droplet soliton,  the length along $z$ direction $l_z$  and cross section in the $x$-$y$ plane $\sigma_{xy}$
of a single droplet as well as the maximum density of molecules $d_{\mathrm{max}}$   in a single droplet increase as the number of molecules $N$ increases.   
 Also, for a fixed $N$,  the dimensions  $l_z$  and $\sigma_{xy}$ and the density  $d_{\mathrm{max}}$ are larger for  a two-droplet soliton than those of a four-droplet soliton.  

From Fig. \ref{fig1} we find that the two-droplet and four-droplet solitons are excited states, whereas the ground state is an one-droplet soliton.  But these states have distinct spatial symmetry properties which attributes remarkable stability properties to these states.  To illustrate this stability further, we investigate the uniform motion of these solitons along $z$ direction, which is the fundamental property of a soliton. To set these solitons in uniform motion along the $z$ direction with a velocity $v$, the converged wave function obtained by imaginary-time propagation is multiplied by a prefactor  $\exp(\mathrm{i} v z)$ 
and the resultant wave function is then used as   the initial wave function in real time propagation.  In the ideal case of infinitely  small space and time steps, the real-time dynamics generates a moving soliton of velocity $v$. But in actual calculation with a finite space and time step, the generated uniform velocity is slightly smaller than $v$.

 \begin{figure}[t!]
\begin{center}
\includegraphics[width=.22\linewidth]{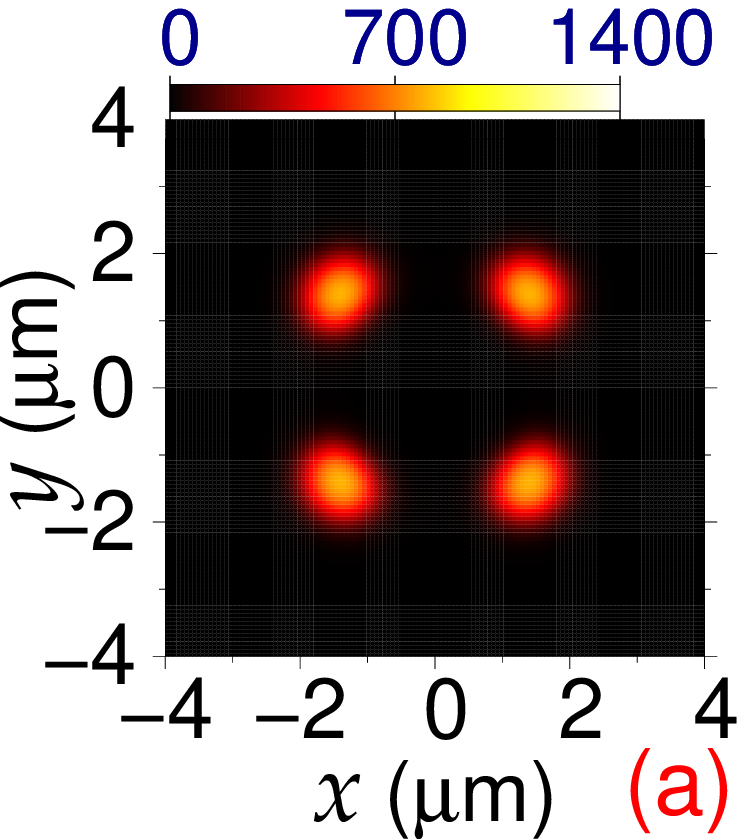} 
\includegraphics[width=.26\linewidth]{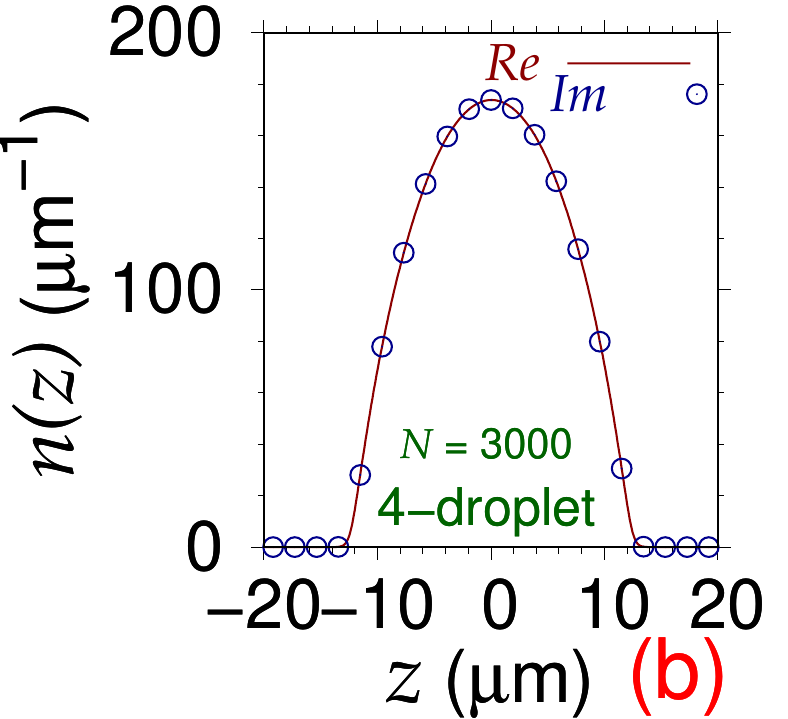}
\includegraphics[width=.48\linewidth]{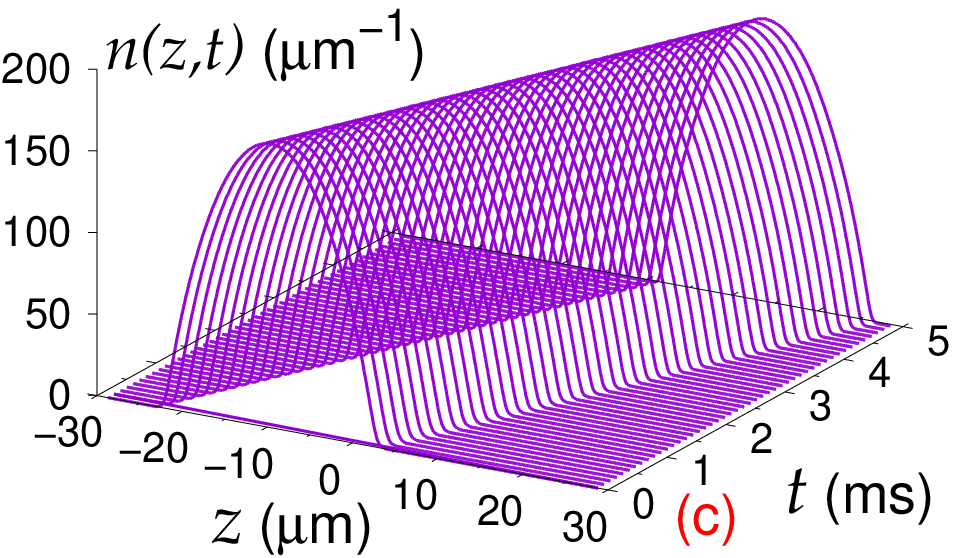}
\includegraphics[width=.22\linewidth]{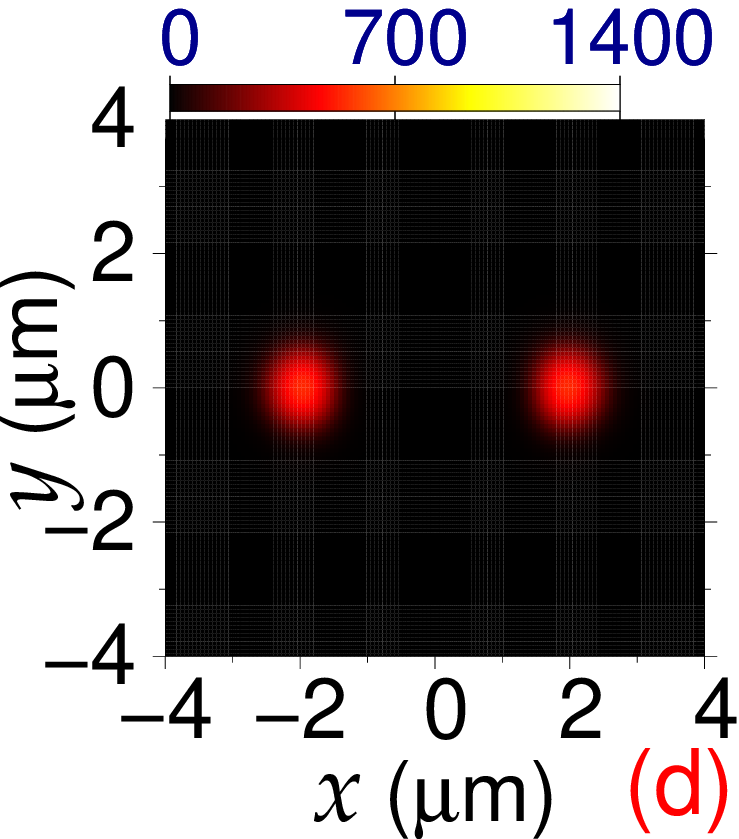} 
\includegraphics[width=.26\linewidth]{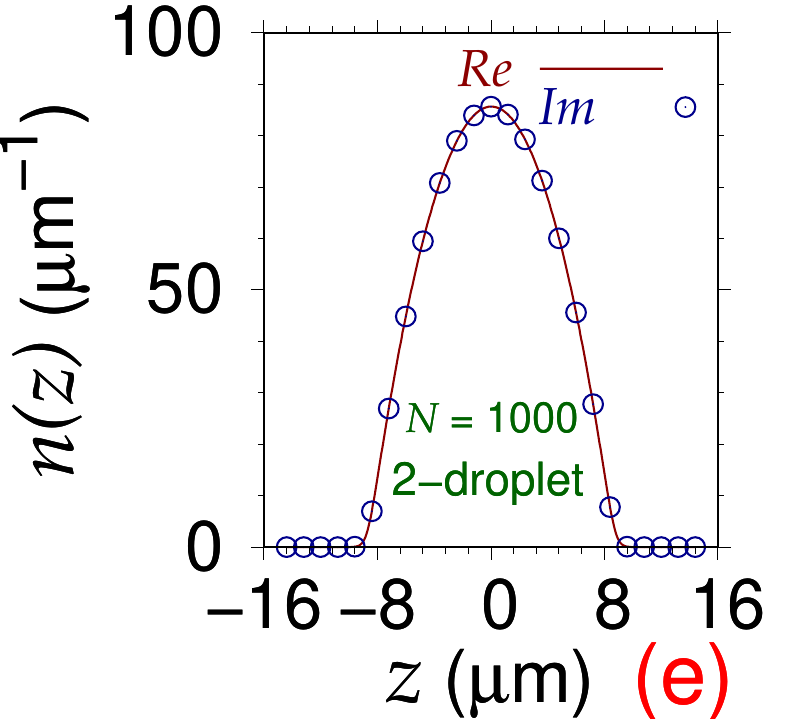}
\includegraphics[width=.48\linewidth]{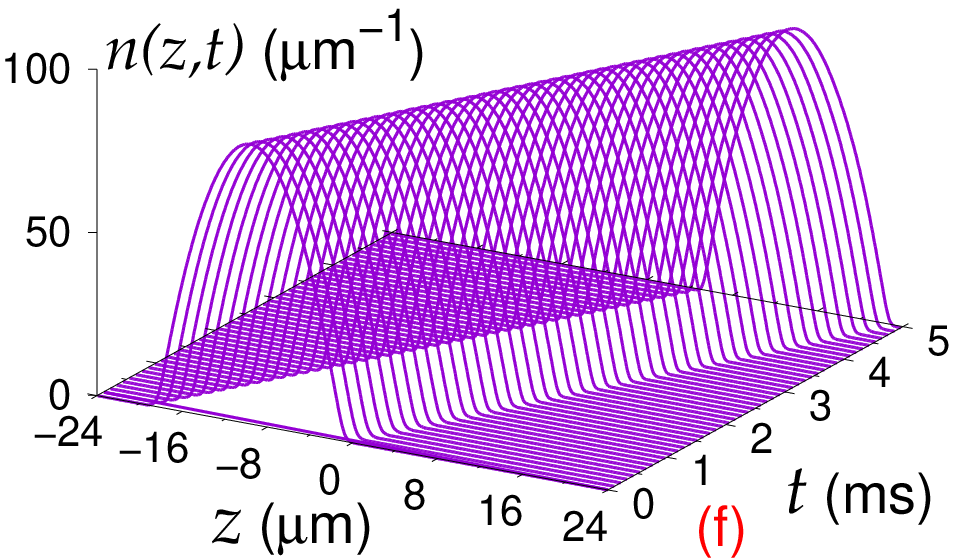}

\caption{Contour plot of (a) net integrated 2D density  $n(x,y)\equiv Nn_{\mathrm{2D}}(x,y)$   and plot of (b) net  integrated 1D density  $n(z)\equiv Nn_{\mathrm{1D}}(z)$ ($Re$) the four-droplet soliton of  $N=3000$ NaCs molecules, viz. Fig. \ref{fig3}(c),   after uniform motion along the $z$ axis during 5 ms. In (b) we also present the initial imaginary-time profile ($Im$).
Both profiles are centralized at $z=0$ in (b).
 (c) Time evolution of net integrated 1D density  $n(z,t)$ during the uniform motion  of the four-droplet soliton. Contour plot of (d) net  integrated 2D density  $n(x,y)$   and plot of (e) net integrated 1D density  $n(z)$ of  the two-droplet soliton of  $N=1000$ NaCs molecules, viz. Fig. \ref{fig3}(e),   after uniform motion along the $z$ axis during 5 ms.  (f) Time evolution of net integrated 1D density  $n(z,t)$ during the uniform motion  of the two-droplet soliton. 
}

\label{fig7} 
\end{center}
\end{figure}
 
To test the dynamical stability  and mobility of the four- and two-droplet solitons, we consider the converged imaginary-time wave functions corresponding to the four-droplet soliton of  $N=3000$ molecules, viz.  Fig. \ref{fig3}(c) and the two-droplet soliton of $N=1000$ molecules, viz. 
 \ref{fig3}(e), respectively,  multiply these by the phase factor $\exp(\mathrm{i}vz)$ and use as the initial function in real-time propagation during an interval of time 5  ms.  
In  Fig. \ref{fig7}(a) we illustrate the contour plot of the net integrated 2D density $n(x,y) \equiv  Nn_{\mathrm{2D}}(x,y)$  versus $x,y$ of the four-droplet soliton of $N=3000$  NaCs molecules  as obtained by real-time propagation at $t=5$ ms. 
In Fig. \ref{fig7}(b) we display the net integrated 1D density $n(z) \equiv Nn_{\mathrm{1D}}(z)$ along $z$ direction  at  $t=5$ ms as obtained by real-time propagation  ($Re$)  and compare the same with the initial integrated 1D density $n(z)$  as obtained by imaginary-time propagation ($Im$). 
We needed to translate the real-time profile by a distance $z=z_0= 23.5$ $\mu$m along the $z$ direction  so that the peak of the  densities of real- and imaginary-time results coincide in Fig. \ref{fig7}(b). Hence the four-droplet soliton has moved a distance of $z_0=23.5$  $\mu$m in 5 ms resulting in a velocity of  $v_0 =4.7$  mm/s. In Fig. \ref{fig7}(c) we show this propagation of the four-droplet soliton through a plot of the net integrated 1D density $Nn(z,t)$ versus $z$ and $t$ during the uniform linear motion along the $z$ axis with the velocity of 4.7 mm/s.  Similarly, in Fig. \ref{fig7}(d)  we present the contour plot of the net integrated 2D density $n(x,y)$ 
of the two-droplet soliton of $N=1000$ NaCs molecules after uniform linear motion during 5 ms. The net integrated 1D density $n(z)$ of the same at  $t=5$ ms ($Re$) is compared with the initial density ($Im$) in Fig. \ref{fig7}(e).    In Fig. \ref{fig7}(f) we present the time evolution of the net integrated 1D density $n(z,t)$  during this linear motion. We attributed the same velocity (4.7 mm/s) and same displacement (23.5 $\mu$m)  of the four- and the two-droplet solitons in Figs. \ref{fig7}(a)-(c) and \ref{fig7}(d)-(f),  respectively. The integrated 1D and 2D profiles $n(z)$  and $n(x,y)$ remain practically unchanged during and after the uniform motion during 5 ms; this demonstrates  the dynamical stability and the mobility of the solitons.

\section{Summary}

\label{IV}  
 
We have demonstrated, using an improved mean-field model including a higher-order LHY repulsive interaction \cite{lhy}, as modified for dipolar systems \cite{qf1,qf2}, in the GP model equation,  that in a very strongly dipolar BEC of NaCs molecules \cite{NaCs} one can have a metastable (excited state) two- and a four-droplet quasi-1D soliton axially free to move along the $z$ direction.   These solitons are subject to  an expulsive Gaussian potential  and a weak harmonic potential
$-$ both in the $x$-$y$ plane. In this study the dipolar and scattering lengths were taken to be $a_{\mathrm{dd}}= 2000a_0$ and $a=100a_0$, and the number of molecules were kept between 1000 and 4000.
 The ground state of this system is an one-droplet soliton.  The two-droplet (four-droplet) soliton is the first (second) excited state as illustrated in Fig. \ref{fig1}.  The droplets are localized in the $x$-$y$ plane and elongated along the $z$ axis, like the droplets in a fully trapped dipolar BEC \cite{y13,2d3}, as shown in Fig. \ref{fig3}.
 The repulsion   among (between) the droplets of the four-droplet (two-droplet) soliton, arising due to the combined action of the expulsive Gaussian potential and the dipolar interaction, stabilizes the formation of the soliton. 
The GP model has a net cubic attractive nonlinear term and the higher-order   LHY interaction leads to a repulsive quartic nonlinearity, which stops the collapse of the strongly dipolar NaCs condensate.
 These  dynamically stable solitons are phase coherent  and hence can move as a unified whole   along the polarization $z$ direction with a constant velocity, without any relative motion between the droplets, as shown by real-time simulation in Fig. \ref{fig7}, employing the converged imaginary-time wave function as the initial state.  To start the motion of the soliton in real-time propagation with a velocity $v$ along $z$ direction, a phase factor of $\exp(izv)$ is printed on the initial state.
In this fashion, a two-droplet soliton  of $N=1000$ NaCs molecules and a four-droplet soliton of $N=3000$ NaCs molecules 
moved 23.5 $\mu$m  in 5 ms with a linear velocity of 4.7 mm/s, viz.  Fig. \ref{fig7},  without any visible deformation of shape  demonstrating the mobility and dynamical stability of the solitons. The four- and the two-droplet solitons have a topology distinct from that of  a usual one-droplet soliton and the very large dipole moment of the NaCs molecules has been fundamental in generating these solitons.  If the net dipolar interaction of the system is reduced, 
keeping all other parameters unchanged, the droplets of the present solitons join together to form a  one-droplet soliton. 

\acknowledgments
   
SKA acknowledges support by the CNPq (Brazil) grant 301324/2019-0.


\end{document}